\documentclass[%
 reprint,
 amsmath,amssymb,
 aps,
floatfix
]{revtex4-2}

\usepackage{hyperref}
\hypersetup{
    colorlinks=true,
    linkcolor=blue,
    filecolor=magenta,      
    urlcolor=blue,
    pdfpagemode=FullScreen,
}

\usepackage{tikz}
\usepackage{quantikz}
\usepackage[caption = false]{subfig}
\usepackage[utf8]{}
\usepackage{float}
\usepackage{graphicx}
\usepackage{dcolumn}
\usepackage{bm}

\begin{document}

\title{Digital Quantum Simulation of Reaction-Diffusion Systems on Lattice}

\author{Louie Hong Yao}
\email{lhyao731@gmail.com}

\date{\today}

\begin{abstract}
The quantum computer offers significant advantages in simulating physical systems, particularly those with exponentially large state spaces, such as quantum systems. Stochastic reaction-diffusion systems, characterized by their stochastic nature, also exhibit exponential growth in the dimension of the state space, posing challenges for simulation at a probability distribution level. We explore the quantum simulation of stochastic reaction-diffusion systems on a digital quantum computer, directly simulating the system at the master equation level. Leveraging a spin representation of the system, we employ Trotterization and probabilistic imaginary time evolution (PITE) to simulate the probability distribution directly. We illustrate this approach through four diverse examples, ranging from simple single-lattice site generation-annihilation processes to a system featuring active-absorbing phase transition.
\end{abstract}

\maketitle

\section{Introduction}
Reaction-diffusion systems depict the dynamics of particles as they diffuse through space and engage in local contact reactions. Despite their apparent simplicity, these systems exhibit a wide array of fascinating phenomena, including dynamical phase transitions\cite{janssen1981nonequilibrium,grassberger1981phase,cardy1998field,antal2001phase,deng2020coupled,howard1997real,jensen1993critical,janssen2004pair,nettuno2024role}, pattern formation \cite{ben2003bifurcations,yao2023perturbative, cross1993pattern,butler2009predator,tauber2011stochastic,dobramysl2018stochastic}, and self-organization \cite{reichenbach2007mobility,reichenbach2008self,frachebourg1996spatial}. Moreover, they serve as an effective framework for describing numerous phenomena across diverse fields such as biology, ecology, chemistry, economics, epidemiology, and game theory \cite{elton1942ten, maynard1974models,cheng2013cortical,durrett1999stochastic,hofbauer1998evolutionary,tauber2012population,tauber2014critical,lipowski1999oscillatory,frey2010evolutionary}. 

Traditionally, reaction-diffusion dynamics have been described using interconnected nonlinear ordinary differential equations. To account for spatially extended systems, this approach is extended to include partial differential equations, incorporating species diffusion through simple diffusion mechanisms. However, this traditional mean-field or mass-action approach overlooks the inherent randomness and stochasticity in these processes. These fluctuations, stemming from discrete individual numbers, and spatio-temporal correlations are disregarded. Yet, these fluctuations and correlations can lead to behaviors significantly different from those predicted by mean-field theory \cite{durrett1999stochastic}.

The nonlinearities inherent in mean-field theory already pose challenges for solving these systems, and the introduction of stochasticity further complicates matters. Despite numerous attempts \cite{henkel2001exact,henkel1997reaction,alcaraz1994reaction,dimentberg2003stochastic}, exact solutions are rarely attainable due to the systems' inherent stochasticity and nonlinearity. Consequently, various analytical and numerical methods have been employed to study them, including field theoretical approaches (see, for example, \cite{tauber2014critical, yao2023perturbative, jensen1993critical, cardy1998field, tauber2012population}), agent-based Monte Carlo simulations (see, for example, \cite{lipowski1999oscillatory,dobramysl2018stochastic, swailem2024computing,deng2020coupled}), and tensor network-related algorithms \cite{carlon1999density,helms2019dynamical,merbis2023efficient,gu2022tensor,kemper2002stochastic,johnson2010dynamical}. However, most of these methods focus on understanding the system at an averaged level by calculating the expectation values of observables of interest. Although the master equation of the system is not difficult to write down, the stochastic nature of the system leads to an exponentially growing state space dimension. Consequently, conducting a probability distribution level study within the framework of the master equation is challenging due to constraints on computational power. Even for the simplest case with only a single species of particles on lattice sites, the dimension of the state space scales as $2^N$ with $N$ as the number of lattice sites. For $N > 100$, simulations become infeasible at all.

Quantum computers, leveraging qubits instead of classical bits, benefit from the exponentially expanding Hilbert space as the number of qubits increases. This unique feature has enabled them to demonstrate advantages in simulating physical systems, especially those with exponentially large numbers of degrees of freedom, such as quantum and stochastic systems. Various algorithms have been devised to tackle these tasks. 
For instance, variational quantum algorithms such as the variational quantum eigensolver \cite{tilly2022variational, peruzzo2014variational, tang2021qubit} and quantum approximate optimization algorithms \cite{farhi2014quantum, zhou2020quantum, sridhar2023adapt} are designed to compute the eigenspectrum of quantum systems through variational techniques. Methods are also proposed to simulate quantum dynamics, in closed system \cite{trabesinger2012quantum, georgescu2014quantum, brown2010using, lloyd1996universal, nielsen2002universal, berry2015simulating, low2017optimal,childs2012hamiltonian}, open systems \cite{wang2011quantum, endo2020variational,sweke2015universal} and in imaginary time for systems in thermal equilibrium \cite{jones2019variational, mcardle2019variational, motta2020determining, sun2021quantum, nishi2021implementation, leadbeater2023non}. 
Algorithms have also been developed to tackle partial differential equations \cite{cao2013quantum, berry2014high, montanaro2016quantum, linden2022quantum}, simulate stochastic systems \cite{kubo2021variational, banchi2023accuracy, ghafari2019dimensional, korzekwa2021quantum, kiss2023importance, blank2021quantum, kalantzis2024quantum}, and explore fluid dynamics \cite{yepez1998quantum, gaitan2020finding, steijl2019quantum}. The successful implementation of these algorithms on real quantum devices (see, for example, \cite{kapil2018quantum, jafferis2022traversable, kamakari2022digital, pagano2020quantum, semeghini2021probing}) addresses the efficacy of quantum computing in simulating large-scale systems.

In this paper, we leverage the power of quantum computers to simulate stochastic reaction-diffusion systems on lattices by directly simulating the master equation governing the system. Our approach begins by mapping the reaction-diffusion systems to spin models \cite{henkel1997reaction, henkel2001exact}, laying the groundwork for direct simulation using qubits. The dynamics of the system are represented by a spin `pseudo'-Hamiltonian or spin Liouvillian $H$, with the time evolution governed by the operator $e^{-Ht}$. Since this `pseudo'-Hamiltonian is generally non-anti-Hermitian, the time evolution is non-unitary. To simulate this operator, we employ Trotterization methods. Additionally, we utilize the probabilistic imaginary time evolution method proposed in \cite{leadbeater2023non} to implement the non-unitary components within the Trotterization framework. To bridge the gap in normalizations between quantum and stochastic states, we introduce specialized pre- and post-processing techniques, eliminating the requirement for exponential classical post-processing. We then put our methods to the test across four diverse examples, spanning from simple single-site generation and annihilation processes to a system demonstrating active-absorbing phase transition. Through these examples, we showcase the effectiveness of our simulation techniques in capturing various dynamic behaviors. While our simulation results demonstrate promise in phases characterized by rapid system relaxation, challenges emerge near critical points where slow relaxation contributes to error accumulation, ultimately impacting result accuracy. In our outlook, we propose potential solutions to address these limitations and further enhance the robustness of the simulation results.

The article is structured as follows: In the subsequent section, we provide an overview of reaction-diffusion dynamics and introduce spin representations of these systems. Section \ref{sect: methods} delves into our simulation methodologies, covering Trotterization, Pauli gadget, probabilistic imaginary time evolution, and our streamlined post-processing approach. We then proceed to apply these simulation techniques to four distinct models in Section \ref{sect: models results}, where we compare quantum simulation results with exact outcomes. Finally, we wrap up with a concise summary and offer insights into future directions.

\section{A Brief Review of Reaction-Diffusion Dynamics and Spin Representations}
Before we discuss the reaction-diffusion dynamics, we first review the master equation description of the stochastic systems and present the pseudo-Hamiltonian representation of the master equations. Given the probability distribution $P$ of the system and the state space $\mathcal{H}$, most if not all markovian stochastic dynamics can be written as
\begin{equation}
\frac{\mathrm{d}}{\mathrm{d}t}P_i(t) = \sum_{j=1}^{D}W_{ij}P_j(t) - \sum_{j= 1}^{D}W_{ji}P_i(t)
\end{equation}
where $D$ is the dimension of the state space and $i = 1,2,3,...,D$ represent the states in the state space $\mathcal{H}$. $W_{ij}$ are the transition rates from state $j$ to state $i$.

Due to the linearity of the master equation, we can define $|\Psi(t)\rangle = \sum_{i}P(i,t)|i\rangle$, where $\{|i\rangle\}$ are orthogonal to each other. The master equation can then be written as a `pseudo'-Schrodinger equation:
\begin{equation}
    \frac{\mathrm{d}}{\mathrm{d}t}|\Psi(t)\rangle = -H|\Psi(t)\rangle,
\label{pschodingereq}
\end{equation}
where the state $|\Psi(t)\rangle = \sum_i P_i(t)|i\rangle$ represents the probability distribution of the system and $H$ is the pseudo-Hamiltonian (or Liouvillian) with
\begin{equation}
    H|j\rangle = -\sum_{i}W_{ij}|i\rangle +\sum_{i}W_{ij}|j\rangle.
\end{equation}
The `pseudo'-Schrodinger equation Eq.(\ref{pschodingereq}) can be formally solved as
\begin{equation}
    |\Psi(t)\rangle = U(t)|\Psi(0)\rangle.
\end{equation}
with the time-evolution operator represented as $U(t) = e^{-Ht}$. Considering observables represented by the operator $O$, the expectation values and two-point correlation functions can be expressed as:
\begin{equation}
    \begin{aligned}
        \langle O(t)\rangle &= \langle P|\hat{O} e^{-Ht}|\Psi(0)\rangle,\\
        \langle O(t_2)O(t_1)\rangle &= \langle P|\hat{O} e^{-H(t_2-t_1)}\hat{O}e^{-H t_1}|\Psi(0)\rangle
    \end{aligned}
\end{equation}
where $\langle P| = (1,\ 1,\ ..., 1)$ is a projection state. The normalization condition of the probability is then given by $\langle P|\Psi \rangle = 1$.

The reaction-diffusion dynamics involve the diffusion of particles and their reactions when they come into contact. A simple example is pair annihilation dynamics, where particles of the same type move around and may vanish when they collide. 
In this paper, we're focusing on systems arranged on lattices with just one species of particles allowed. Our simulation methods can be extended to systems with multiple species, either by using qudits or by using multiple qubits to represent one lattice site.
We also enforce a rule that each lattice site can only hold one particle, which is typical in Monte Carlo simulations for these systems.
Each lattice site then has two possible states: occupied (denoted as $|\uparrow\rangle$) or empty (denoted as $|\downarrow\rangle$), making the state space for a single site $\mathcal{H}_2 = \mathbb{R}^2$.
Extending this concept to a lattice comprising $N$ sites yields the comprehensive state space $\mathcal{H} = \mathcal{H}_2^{\otimes N}$.
Particle generation and annihilation are succinctly captured by the operators $\sigma^\pm = (\sigma^x \pm i\sigma ^y)/2$, following established conventions of quantum mechanics.
Hence, particle interactions can be depicted by utilizing these operators.
A fundamental instance arises in the annihilation of a single particle with a rate $\lambda$. The Hamiltonian governing this single-particle reaction is depicted as $\lambda\sum_i(n_i - \sigma_i^-)$, where $n_i = \sigma_i^+\sigma_i^- = (\sigma_i^z + 1)/2$ represents the particle number operator, and the index $i$ denotes the lattice sites. Other reactions can also be expressed in this way. We outline the Hamiltonians corresponding to common reactions in Table \ref{table: pseudo hamiltonians}. In cases involving multiple reactions, the Hamiltonian for the entire system is derived by summing the Hamiltonians of each reaction. For example, the system with particle diffusion and pair annihilation can be represented by
\begin{equation}
\begin{aligned}
    H = &D\sum_{\langle i,j\rangle}\Bigg[(\sigma_i^+-\sigma_j^+)(\sigma_i^- - \sigma_j^-) - 2n_in_j\Bigg]\\
    &+ \lambda\sum_{\langle i,j\rangle}(n_in_j - \sigma_i^-\sigma_j^-),
\end{aligned}
\end{equation}
where $\langle i,j\rangle$ denotes nearest neighbors.
Given that the three Pauli matrices $\sigma^x,\ \sigma^y,\ \sigma^z$ and the 2-dimensional identity matrix constitute a complete basis for $SL(2,\mathbb{C})$, the Hamiltonian can be represented as linear combinations of tensor products of the Pauli basis $\{1,\sigma^x,\sigma^y, \sigma^z\}$. The expression of the Hamiltonian in the Pauli basis is also included, as exponentials of Pauli matrices are simpler to implement on a digital quantum computer.
\begin{widetext}

\begin{table}
\begin{tabular}{c | c c} 
 \hline
     & \multicolumn{2}{c}{Pseudo-Hamiltonians} \\ [0.5ex] 

 Reactions & Generation-Annihilation basis & Pauli basis\\
 \hline\hline
 Free particles Hopping & $
     \sum_{\langle i,j\rangle}(\sigma_i^+-\sigma_j^+)(\sigma_i^- - \sigma_j^-) - 2n_in_j$ & $-\frac{1}{2}\sum_{\langle i,j\rangle} \Big(\sigma_i^x\sigma_j^x + \sigma_i^y\sigma_j^y + \sigma_i^z\sigma_j^z - 1\Big)$ \\ 
 \hline
Pair Annihilation ($2A\rightarrow 0$) & $\sum_{\langle i,j\rangle}(n_in_j - \sigma_i^-\sigma_j^-)$ & $\begin{aligned}\frac{1}{4}\sum_{\langle i,j\rangle}\Big(-\sigma_i^x\sigma_j^x + \sigma_i^y\sigma_j^y + \sigma_i^z\sigma_j^z +i\sigma_i^x\sigma_j^y + i\sigma_i^y\sigma_j^x\Big)\\ + \frac{1}{2}\sum_i\left(\sigma_i^z+\frac{1}{2}\right)\end{aligned}$ \\
 \hline
  Pair Coagulation ($2A\rightarrow A$)& $\sum_{\langle i,j\rangle}n_in_j - \frac{1}{2} n_i \sigma_j^- - \frac{1}{2}n_j \sigma_i^-$ & $\begin{aligned}\frac{1}{8}\sum_{\langle i,j\rangle}\Big(2\sigma_i^z\sigma_j^z - \sigma_i^z\sigma_j^x - \sigma_i^x\sigma_j^z +i\sigma_i^z\sigma_j^y + i\sigma_i^y\sigma_j^z\Big)\\ + \frac{1}{4}\sum_i\left(2\sigma_i^z-\sigma_i^x + i\sigma_i^y + \frac{1}{4}\right)\end{aligned}$ \\
 \hline
Spontaneous decay ($A\rightarrow 0$)& $\sum_{i}n_i - \sigma_i^-$ & $\frac{1}{2}\sum_{i}\Big(-\sigma_i^x + i\sigma_i^y + \sigma_i^z + 1\Big)$ \\
 \hline
 Spontaneous generation ($0\rightarrow A$)& $\sum_{i}(1 - n_i) - \sigma_i^+$ & $\frac{1}{2}\sum_i\Big(1-\sigma_i^z - \sigma_i^x -i\sigma_i^y\Big)$ \\

 \hline
Branching ($A\rightarrow 2A$) & $\sum_{\langle i,j\rangle}n_i + n_j - 2n_in_j - n_i\sigma_j^+ - n_j\sigma_i^+   $ & $\begin{aligned}-\frac{1}{4}\sum_{\langle i,j\rangle}\Big(2\sigma_i^z\sigma_j^z + \sigma_i^z\sigma_j^x + \sigma_i^x\sigma_j^z +i\sigma_i^z\sigma_j^y + i\sigma_i^y\sigma_j^z\Big)\\ - \frac{1}{2}\sum_i\left(\sigma_i^x + i\sigma_i^y - 1\right)\end{aligned}$  \\ [1ex] 
 \hline
\end{tabular}
    \caption{The Pseudo-Hamiltonians for common reactions involving particle species $A$ are detailed above. Expressions are provided both in terms of $\{\sigma^+,\sigma^-, n\}$ and $\{1, \sigma^x,\sigma^y, \sigma^z\}$. As we only consider contact reactions and nearest-neighbor hopping, $\langle i,j\rangle$ represents nearest-neighbor pairs.
    }
    \label{table: pseudo hamiltonians}
\end{table}

\end{widetext}

\section{Methods}
\label{sect: methods}
In this section, we delve into the simulation methods employed to simulate the reaction-diffusion system on a digital quantum computer. Additionally, we elaborate on several implementation details pivotal to our simulation process.

\subsection{Trotterization}
Quantum simulation of dynamics, whether in real-time, imaginary time, or involving general open quantum dynamics, requires implementing operators of the form $e^{-Ht}$, where $H$ represents the system's Hamiltonian, Liouvillian, or Lindbladian. Even if $H$ consists solely of local terms, i.e., $H = \sum_i H_i$ where $H_i$s are local, the operator $e^{-Ht}$ can still pose significant nonlocality challenges for implementation. Various simulation methods have been proposed to tackle this challenge, including the Lie-Trotter-Suzuki decomposition \cite{lloyd1996universal, nielsen2002universal}, linear combination of unitaries (LCU) \cite{berry2015simulating, childs2012hamiltonian, berry2014exponential}, and quantum signal processing (QSP) \cite{low2017optimal}. In this paper, we choose trotterization due to its intuitive nature, ease of implementation, and demonstrated good empirical performance \cite{Childs_2018}.

Given the challenge of handling the operator $e^{-Ht}$, our aim is to decompose it into a product $\prod_i e^{-H_it}$, where $e^{-H_i t}$ terms are more manageable. In many physical systems, including the reaction-diffusion systems under discussion, $H$ can naturally be expressed as $H = \sum_i H_i$, where $H_i$s represent local terms and are easy to implement.

However, due to the non-commutativity between $H_i$s, $e^{-\sum_i H_i t}$ isn't strictly equal to $\prod_i e^{-H_i t}$. An effective solution is offered by the Lie product formula, where given two arbitrary matrices $A$ and $B$, we have:
\begin{equation}
e^{A+B} = \lim_{n\rightarrow\infty}(e^{\frac{A}{n}}e^{\frac{B}{n}})^n.
\end{equation}
Thus, we can divide $e^{-H t}$ into small trotter steps $\Delta t$ and express it as:
\begin{equation}
e^{-Ht} = \lim_{\Delta t\rightarrow 0}\left( \prod_i e^{-H_i \Delta t}\right)^{\frac{t}{\Delta t}}.
\end{equation}
This concept of trotterization has been utilized by physicists for some time, including in the derivation of coherent state path integrals \cite{wen2004quantum, tauber2014critical}. However, in the implementation on a quantum device, we cannot take the limit $\Delta t\rightarrow \infty$. It has been shown that, by the Baker–Campbell–Hausdorff formula, to first order \cite{lloyd1996universal}:
\begin{equation}
e^{-\sum_i H_i t} = \left( \prod_i e^{-H_i \Delta t}\right)^{\frac{t}{\Delta t}} + \mathcal{O}\left(\frac{t^2}{n}\right),
\end{equation}
where $n = t/\Delta t$. 
Thus, by choosing small enough $\Delta t$, we can approximate the real dynamics $e^{-Ht}$ very effectively. 
A more detailed error analysis of this method can be found in \cite{childs2021theory}. 
Throughout this paper, we will only consider the first-order approximation. 

Other methods exist for decomposing $e^{-Ht}$ exactly, such as the Cartan decomposition \cite{kokcu2022fixed}. However, these methods require extensive classical preprocessing and we will not utilize them in this work.

\subsection{Pauli Gadgets}
\label{sect: pauli gadgets}
The Hamiltonian $H$ of spin systems and reaction-diffusion systems with site-restriction can be expressed as a linear combination of tensor products of Pauli matrices. This implies that the $H_i$s are tensor products of Pauli matrices. Implementing these operations on quantum computers necessitates expressing these gates using universal gate sets, typically Clifford gates with a single non-Clifford gate. One straightforward method is through the use of Pauli gadgets, which initially transform the operation to the $z$-basis via rotations in single-qubit Hilbert space. Then, information is propagated to a single target qubit, where a special single-qubit rotation is applied.
For example, consider the operation $e^{-i\sigma^x\otimes\sigma^y\otimes \sigma^zt}$. We can express it as:
\begin{equation}
e^{-i\sigma^x\otimes\sigma^y t} = (H\otimes S)(H \otimes I)e^{-i\sigma^z\otimes \sigma^z t} (H\otimes H)( S^\dagger \otimes I),
\end{equation}
where
\begin{equation}
H = \frac{1}{\sqrt{2}}\begin{pmatrix}
1 & 1 \\
1 & -1
\end{pmatrix},\qquad
S = \begin{pmatrix}
1 & 0 \\
0 & i
\end{pmatrix}
\end{equation}
are the Hadamard gate and the $S$ gate respectively.
With this transformation, we then only need to implement $e^{-i\sigma^z\otimes \sigma^z \otimes \sigma^z t}$.

\begin{figure}[t]
\resizebox{\columnwidth}{!}{%
\begin{quantikz}[node distance=1.5pt]
\qw &\gate{H}& &\ctrl{1} & & & &\ctrl{1} &&\gate{H}&\qw\\
\qw &\gate{S^\dagger} &\gate{H}&\targ{}&\ctrl{1} & & \ctrl{1} &\targ{}  & \gate{H} & \gate{S}&\qw\\
\qw & & & & \targ{} & \gate{R_Z(2t)} & \targ{} & & & &\qw
\end{quantikz}
}

\caption{Implementing the time evolution operator $e^{-i\sigma^x\otimes\sigma^y\otimes \sigma^zt}$ through Pauli gadgets involves a sequence of steps. Initially, we rotate the system to the Pauli $z$ basis. Subsequently, we propagate information to the third qubit and apply an $R_z$ gate. Finally, we perform the inverse operation to return the system to its original basis.}
\label{fig:pauli gadget}
\end{figure}
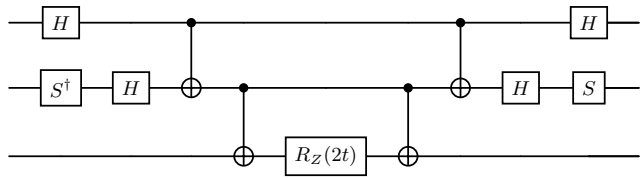

Operations $T\left((\sigma^z)^{\otimes n}\right)$, composed solely of tensor products of Pauli Z matrices, can be expressed as $\sum_i T(p_i)|i\rangle\langle i|$, where $i$ represents binary strings. Here, $p_i$ equals $1$ when the binary string $i$ contains an even number of $1$s, and $p_i$ equals $-1$ when the binary string $i$ contains an odd number of $1$s. We then utilize a CNOT gate to propagate the information to a single qubit, implementing an $R_z$ gate based on the state (0 or 1) of that qubit.
For the example $e^{-i\sigma^x\otimes\sigma^y\otimes \sigma^zt}$, we propagate to the third qubit. The circuit implementation is depicted in Fig. \ref{fig:pauli gadget}.
As plotted in the circuit, we initially employ the $H$ and $S$ gates to transform to the $z$-basis. Subsequently, we utilize CNOT gates to transmit information to the third qubit and apply an $R_z$ gate, i.e.
\begin{equation}
    R_z(\theta) = e^{-i\frac{\theta}{2}\sigma^z} = \begin{pmatrix}
        e^{-i\frac{\theta}{2}} & 0\\
        0 & e^{i\frac{\theta}{2}}
    \end{pmatrix}.
\end{equation}
Following these steps, we revert the system back via inverse transformations. It's essential to note that quantum circuits are applied from left to right, contrary to the usual Dirac notation representation.

In this entire implementation, the Hadamard gate, the $S$ gate, and the CNOT gate are all Clifford gates. The entire procedure only necessitates a single non-Clifford gate, $R_z$. Since Clifford gates are simpler to implement on most current quantum devices, the Pauli gadgets method significantly simplifies the implementation process at the hardware level. Further optimizations of the circuit implementations via Pauli gadgets has also been done and can be found in, for example \cite{van2020circuit, mukhopadhyay2023synthesizing}.

Employing Pauli gadgets offers another advantage: the reduction in the number of ancilla qubits required for non-unitary operations \cite{leadbeater2023non}. During the Pauli gadgets process, information is propagated to a single qubit, necessitating only a single-qubit non-unitary operation. It is well-known that we can purify a system with Hilbert space $\mathcal{H}$ in a tensor product Hilbert space $\mathcal{H}\otimes\mathcal{H}$. Consequently, only a single ancilla qubit is needed to implement the non-unitary operation of the form $e^{-Ht}$ when $H$ is a tensor product of Pauli matrices.

\subsection{Probabilistic Imaginary Time Evolution}
Since the pseudo-Hamiltonian $H$ is not generally anti-Hermitian, simulating the time evolution $e^{-Ht}$ requires implementing non-unitary operations. According to Trotterization to first order, $e^{-Ht}$ can be decomposed into $(\prod_i e^{-H_i t/N})^N$, where $H_i$ can be either Hermitian or anti-Hermitian. For anti-Hermitian $H_i$, $e^{-H_it/N}$ is unitary and thus straightforward to implement. Conversely, for Hermitian $H_i$, implementing $e^{-H_it/N}$ is referred to as imaginary time evolution, as it is equivalent to $e^{-i H_i \tau}$ with $\tau = -it$. The time $t$ here can also be understood as the effective temperature $\beta$ as the imaginary time evolution corresponds to the equilibrium partition function of Gibbs distribution.

Several methods have been proposed for simulating imaginary time evolution on quantum computers. Variational approaches, such as variational imaginary time evolution (VITE) \cite{jones2019variational, mcardle2019variational} and quantum imaginary time evolution (QITE) \cite{motta2020determining, sun2021quantum, nishi2021implementation}, offer a hybrid approach by optimizing a variational ansatz to approximate the target state as closely as possible. Another approach, the probabilistic approach \cite{gingrich2004non, terashima2005nonunitary, nishi2022acceleration,liu2021probabilistic, lin2021real, silva2023fragmented, leadbeater2023non}, which appears more natural to physicists, involves introducing ancilla qubits or environments to the system and then applying post-selection while measuring out the ancilla qubits. In this work, we adapt the methods proposed in \cite{leadbeater2023non}, which leverage the power of the Pauli gadgets introduced in Section \ref{sect: pauli gadgets} to reduce the number of ancilla qubits used in the system.

\begin{figure}[t]
\centering
\subfloat[]{
\begin{quantikz}[node distance=1.5pt,wire
types={q,n},column sep=10pt, row sep={20pt,between origins}]
&& &\qw&\ctrl{1}&\qw & &\\
&&\lstick{$\ket{0}$} &  \setwiretype{q} &\gate{R_x(\theta)}& &\meterD{0\vphantom{0}}&\setwiretype{n}&
\end{quantikz}
}\\ 
\centering
\subfloat[]{
\begin{quantikz}[node distance=1.5pt,wire
types={q,n},column sep=10pt, row sep={20pt,between origins}]
&& &\gate{X}&\ctrl{1}& \gate{X}& &\\
&&\lstick{$\ket{0}$} &  \setwiretype{q} &\gate{R_x(\theta)}& &\meterD{0\vphantom{0}}&\setwiretype{n}&
\end{quantikz}
}
\caption{The quantum circuit implementation of non-unitary operation $e^{-\alpha\sigma^z}$ for (a) $\alpha<0$ and (b) $\alpha > 0$. The angle $\theta$ is given by $\theta = 2\arccos\left(\exp(-2|\alpha|)\right)$.}
\label{fig: implementation of non unitary}
\end{figure}
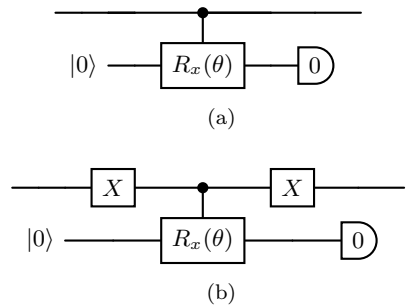

With the Pauli gadgets, the Trotterized real-time evolution operator can be realized by single-qubit gates $e^{-i\alpha\sigma^z}$ with single-qubit rotations and two-qubit control gates, as shown in Fig.\ref{fig:pauli gadget}. Similarly, for the imaginary time evolution operator, we can map it to a single-qubit non-unitary operation $e^{-\alpha \sigma^z}$. Thus, we only need to implement the single-qubit non-unitary operation of the form $e^{-\alpha \sigma^z}$. To do so, we can introduce a single ancilla qubit and implement a controlled-RX gate ($CR_x$). The $R_x$ gate is given by
\begin{equation}
R_x(\theta) = e^{-i\frac{\theta}{2}\sigma^x} = \begin{pmatrix}
\cos\frac{\theta}{2} & -i\sin\frac{\theta}{2} \\
-i\sin\frac{\theta}{2} & \cos\frac{\theta}{2}
\end{pmatrix}.
\end{equation}
Given a state $a|0\rangle + b|1\rangle$ and an ancilla qubit at $|0\rangle$, the $CR_x$ gate transforms the two-qubit state to
\begin{equation}
CR_x\left(a|00\rangle + b|10\rangle\right) = a|00\rangle + b\cos\frac{\theta}{2}|10\rangle -ib\sin\frac{\theta}{2}|11\rangle,
\end{equation}
where the second qubit is the ancilla qubit. By measuring the second qubit and post-selecting it to be $0$, we obtain the final state $a|0\rangle + b\cos\frac{\theta}{2}|1\rangle$ up to a normalization factor. Since the non-unitary operation $e^{-\alpha\sigma^z}$ yields
\begin{equation}
e^{-\alpha\sigma^z}(a|0\rangle + b|1\rangle) \propto a|0\rangle + e^{2\alpha}b|1\rangle,
\end{equation}
when $\alpha<0$, we can implement the $CR_x$ gate with $\theta = 2\arccos\left(\exp(-2|\alpha|)\right)$ and post-select the ancilla qubit to be $0$. This procedure is depicted by the circuit shown in Fig.\ref{fig: implementation of non unitary}(a). When $\alpha > 0$, $e^{2\alpha}$ is larger than $1$ and cannot be mapped to a cosine. Instead, we can use the Pauli $X$ gate to bring $\cos\frac{\theta}{2}$ to the state $|0\rangle$, i.e.,
\begin{equation}
\begin{aligned}
(\sigma^x \otimes 1)CR_x &(\sigma^x \otimes 1)\left(a|00\rangle + b|10\rangle\right) \\ &= a\cos\frac{\theta}{2}|00\rangle + b|10\rangle -ia\sin\frac{\theta}{2}|01\rangle.
\end{aligned}
\end{equation}
By selecting $\theta = 2\arccos\left(\exp(-2|\alpha|)\right)$, we obtain the correct outcome. The circuit for $\alpha > 0$ is shown in Fig. \ref{fig: implementation of non unitary}(b). With Trotterization and Pauli gadgets combined, we are now equipped to implement any time evolutions with the 'Pseudo'-Hamiltonian listed in Table \ref{table: pseudo hamiltonians}.

\subsection{Encoding and Post-processing}

While the pseudo-Schrödinger Eq.(\ref{pschodingereq}) can be represented by combined unitary and non-unitary gates in a quantum circuit, there are still distinctions between this formulation and the quantum-mechanical Schrödinger's equation. For clarity, we denote the quantum states in lowercase and the probability state in uppercase.

One notable difference lies in the normalization of states. In quantum mechanics, state vectors $|\psi\rangle$ are $L_2$ functions with the $2$-norm normalized to $1$, meaning $\langle\psi|\psi\rangle = 1$. However, in the study of reaction-diffusion systems, states are normalized as $\langle P|\Psi\rangle = 1$, or $\sum_i P_i = 1$, in accordance with probability conservation. Consequently, necessary encoding and decoding operations must be performed.

For encoding, we can simply rescale the initial state as follows:
\begin{equation}
|\Psi\rangle = \sum_i P_i|i\rangle \quad \Longrightarrow \quad |\psi\rangle = \frac{1}{\sqrt{\sum_i P_i^2}}\sum_i P_i|i\rangle.
\end{equation}
After the time-evolution, the resulting state from the quantum circuit is expressed as:
\begin{equation}
|\psi_t\rangle = \frac{1}{\sqrt{\langle\psi| U^\dagger_t U_t |\psi\rangle}}U_t|\psi\rangle,
\end{equation}
where $U_t = e^{-Ht}$ represents the time evolution operator.
To ensure correct normalization, we need to compute $\langle P|\psi_t\rangle$. The projection state $\langle P|$ can be represented as:
\begin{equation}
\langle P| = \sqrt{2^n}\langle \mathbf{0}|H^{\otimes n},
\end{equation}
assuming there are $n$ qubits in the system. Here $\langle\mathbf{0}|$ signifies the all-zero state $\langle 0|^{\otimes n}$. As a result, the normalization factor $\langle P|\psi_t\rangle$ is given by:
\begin{equation}
\langle P|\psi_t\rangle = \sqrt{2^n}\langle \mathbf{0}|H^{\otimes n}|\psi_t\rangle.
\end{equation}
By measuring $\langle \mathbf{0}|H^{\otimes n}|\psi_t\rangle$, we obtain the probability amplitude of observing all qubits in the state $H^{\otimes n}|\psi_t\rangle$ to be in the state $|0\rangle$. Essentially, this quantity gives us the amplitude for observing all zeroes when measuring the state $H^{\otimes n}|\psi_t\rangle$. Since this measurement outcome can be directly obtained from the final measurement of the quantum circuit, it eliminates the need for any normalization calculations on a classical computer, which could otherwise be computationally expensive due to its exponential nature.

Since both the normalization factor $\langle P|\psi_t\rangle$ and all components of the state $|\psi_t\rangle$ are real-valued, they can be obtained from measurements of probabilities alone. Consequently, there's no need for additional tomography or complex calculations; everything can be directly inferred from the experimental outcomes.

\section{Models and Results}
\label{sect: models results}
In this section, we apply the methods introduced in Section \ref{sect: methods} to four distinct examples on a one-dimensional lattice. These examples span from a straightforward scenario involving single particle generation and annihilation to a model showcasing an active-absorbing phase transition. Given the limitations of current quantum devices, we perform classical simulations of the process and compare the outcomes with the exact results.

\subsection{Single Particle Generation and Annihilation}
We begin with a simple example involving only a single lattice site. Here, we consider single particle generation and annihilation reactions on the site with rates $\nu$ and $\lambda$, respectively. The `pseudo'-Hamiltonian for this system is given by
\begin{equation}
H = \frac{1}{2}\left[-(\lambda+\nu)\sigma^x + (\lambda-\nu) \sigma^z + i(\lambda-\nu)\sigma^y\right],
\end{equation}
where we have omitted the lower indices for the lattice site. Additionally, we have omitted the constant terms since they only contribute to rescaling in the time-evolution. Neglecting these constants should not affect the results, as we will need to rescale the outcome from the quantum simulation at the end of the process. We will automatically disregard these constant terms in the subsequent calculations unless explicitly stated otherwise. In this system, particle generation competes with annihilation. In the limit $\nu= 0$, the stationary state will be an empty site. Conversely, in the limit where $\lambda = 0$, the particle density of the stationary state will be $1$. For values between these two limits, we will observe a finite particle density $0<\langle n\rangle<1$ in the stationary state.

\begin{figure}[t]
\centering
\subfloat[]{\includegraphics[scale=0.5]{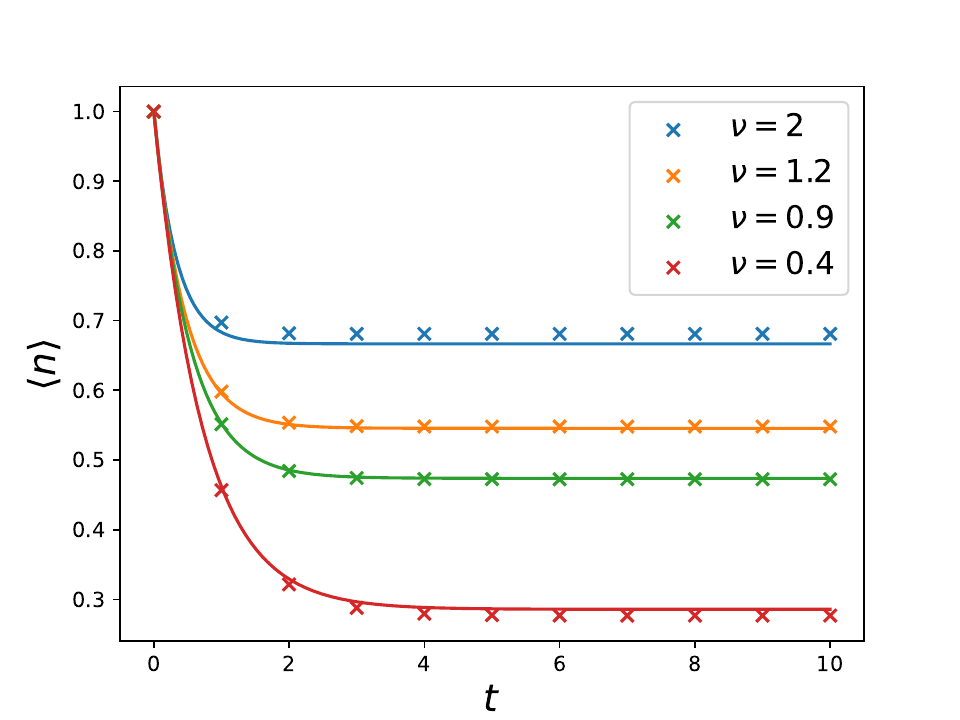}}\\
\centering
\subfloat[]{\includegraphics[scale=0.5]{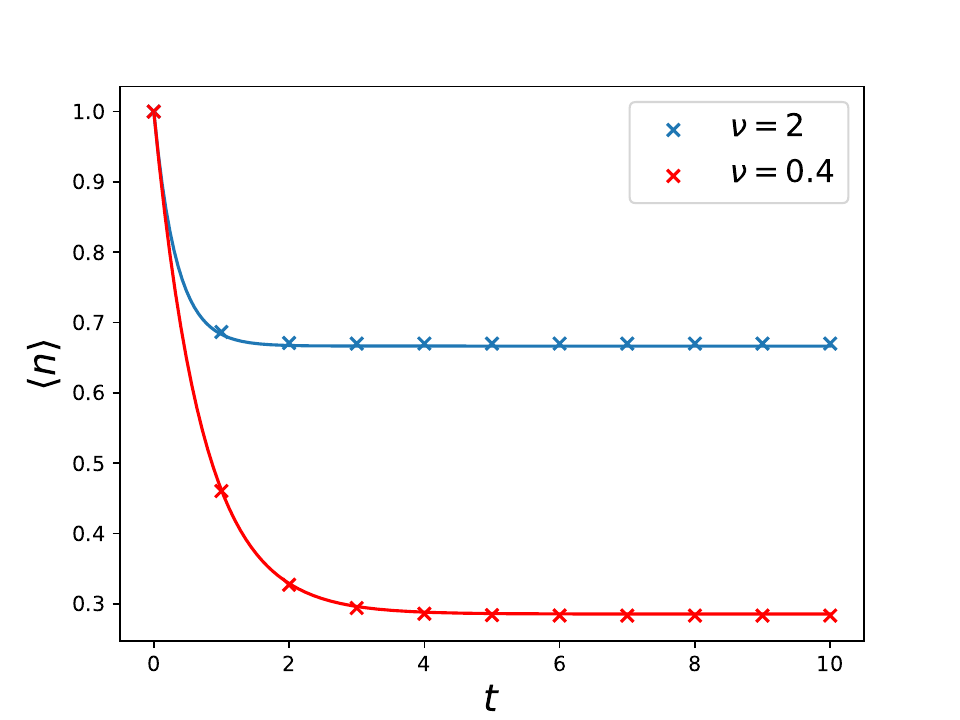}}
\caption{The dynamics of particle density in the single-site generation and annihilation system with fixed generation rate $\lambda = 1$ and varying annihilation rate $\nu$s: In (a), the trotter step size is $\Delta t = 1/20$, while in (b), it is $\Delta t = 1/80$. The solid lines represent the exact results, while the markers indicate the simulated results.}
\label{fig:one site density}
\end{figure}

We simulate the process and depict the time evolution of the particle density $\langle n \rangle$ starting from an occupied state, as shown in Fig.(\ref{fig:one site density}). In the simulation, the 'pseudo'-Hamiltonian is trotterized into terms $e^{(\lambda+\nu)\sigma^x \Delta t}$, $e^{-(\lambda-\nu) \sigma^z\Delta t}$, and $e^{-i(\lambda-\nu)\sigma^y\Delta t}$, where the first two terms are non-unitary and the last term is unitary and can be implemented directly. For simplicity, we choose $\lambda = 1$ in the simulation and compare the results for different values of $\nu$.

In Fig. \ref{fig:one site density}(a), we plot the particle density $\langle n\rangle$ versus time $t$ for different values of $\nu$ with a trotter step size $\Delta t = 1/20$. The solid lines represent the exact results, while the scattered labels represent the results from quantum simulations. We observe that for intermediate values of $\nu$, the process is very well captured by the simulation. However, for both large and small values of $\nu$, there are visible errors, which arise from the fact that either $\lambda + \nu$ or $\lambda-\nu$ is large, leading to increased trotter errors.

We further simulate the cases of large and small $\nu$ values with a smaller trotter step size $\Delta t = 1/50$ in Fig.\ref{fig:one site density}(b). As we decrease the trotter step size, the errors decrease, and the results better agree with the exact results.

\subsection{Free Particles Hopping}
\begin{figure}[t]
\centering
\subfloat[]{\includegraphics[scale=0.5]{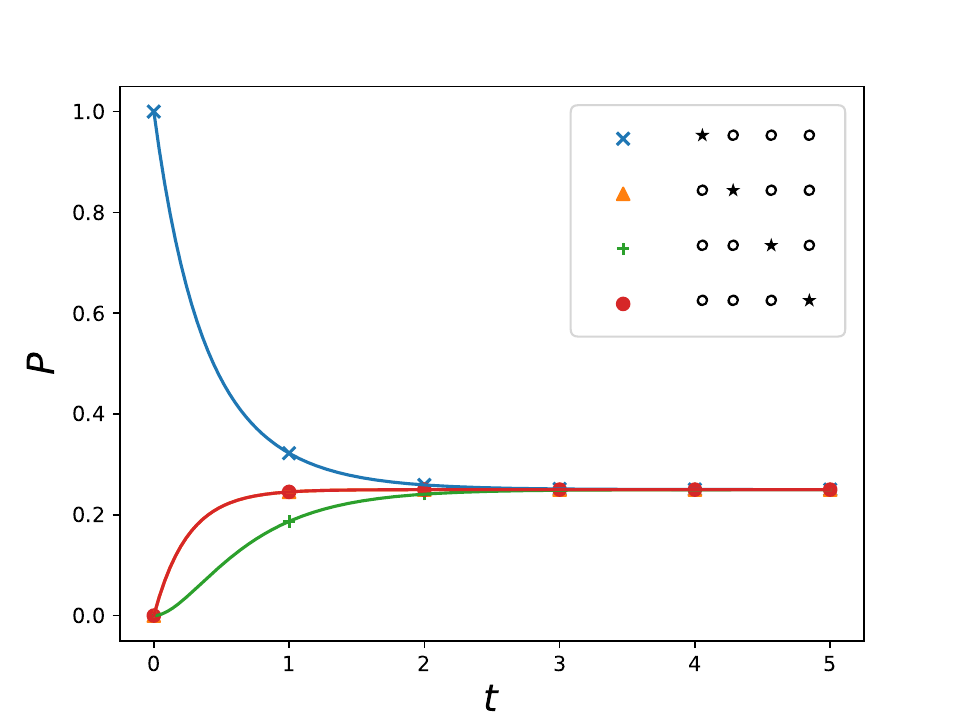}}\\ 
\centering
\subfloat[]{\includegraphics[scale=0.5]{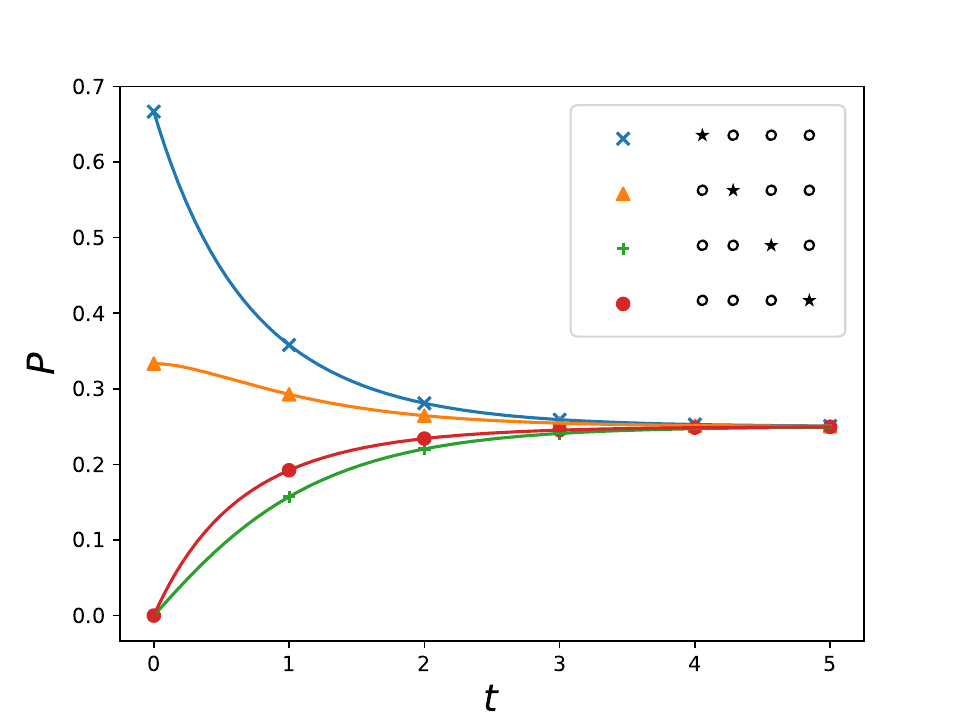}}
\caption{The probabilities associated with different states in the pure hopping dynamics on four lattice sites: In (a), the initial state is $|\bigstar\bigcirc\bigcirc\bigcirc\rangle$, and the diffusion constant is $D = 1$. In (b), the initial state is $2/3|\bigstar\bigcirc\bigcirc\bigcirc\rangle + 1/3|\bigcirc\bigstar\bigcirc\bigcirc\rangle$, and the diffusion constant is $D = 0.6$. Here $\bigstar$ represents the occupied site and $\bigcirc$ represents the empty site.}
\label{fig: pure hopping}
\end{figure}

The next example we consider involves a free particle hopping on a line. Given the diffusion constant $D$, the `pseudo'-Hamiltonian is written as
\begin{equation}
H = -\frac{1}{2}D\sum_{\langle i,j\rangle}\left(\sigma_i^x\sigma_j^x + \sigma_i^y\sigma_j^y + \sigma_i^z\sigma_j^z\right).
\end{equation}
This `pseudo'-Hamiltonian coincides with the Hamiltonian of the Heisenberg model; however, the time evolution is not unitary. We benchmark the simulation on a system with $4$ lattice sites, and the results are plotted in Fig.\ref{fig: pure hopping}. Since particle hopping conserves the total particle number in the system, we instead plot the dynamics of the probability of the relevant states. The trotter step size is chosen as $\Delta t = 1/20$.

In Fig.\ref{fig: pure hopping}(a), we simulate the system starting with the state $\bigstar\bigcirc\bigcirc\bigcirc$ with a diffusion constant $D= 1$. Here, $\bigstar$ indicates the site is occupied, and $\bigcirc$ indicates the site is empty. Since there is only one single particle in the system, only four states are relevant in the dynamics, as shown in the plot. As expected, we observe that the probability of all four states converges to $1/4$. Additionally, due to the periodic boundary condition we choose, the states labeled with a triangle and a plus sign are symmetric to each other. Consequently, their probabilities are equal, leading to an overlap in the plot. 
In Fig.\ref{fig: pure hopping}(b), we present the results from a different initial state: $2/3$ probability of staying in $\bigstar\bigcirc\bigcirc\bigcirc$ and $1/3$ probability of staying in $\bigcirc\bigstar\bigcirc\bigcirc$. Here, we employ a different diffusion constant $D = 0.6$ in the simulation. As anticipated, we observe that the probability of all four relevant states converges to $1/4$, resulting in a uniform distribution. This uniform distribution is in line with the expected behavior of a system undergoing free particle hopping, where the particle density tends to spread out evenly across the lattice over time. 
From the plots, we can see that at a trotter step size $\Delta t = 1/20$, the simulation recovers the exact result very well at the probability density level. This indicates that the trotterization method, combined with the Pauli gadgets, effectively captures the non-unitary evolution described by the 'pseudo'-Hamiltonian.

\subsection{Pair Annihilation}

\begin{figure}[t]
\centering
\subfloat[]{\includegraphics[scale=0.5]{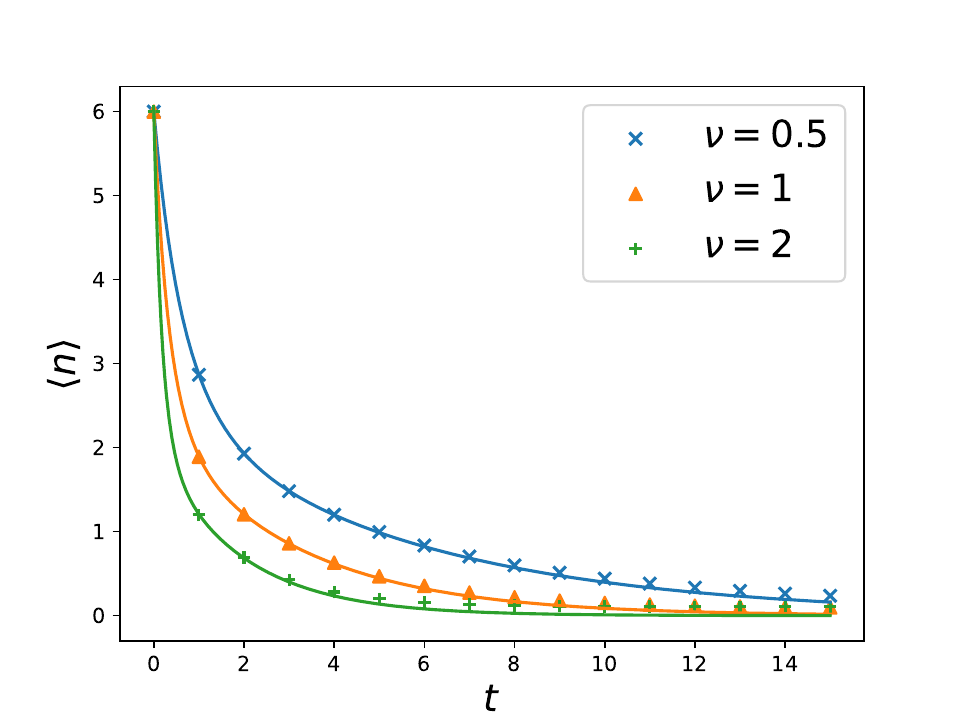}}\\
\centering
\subfloat[]{\includegraphics[scale=0.5]{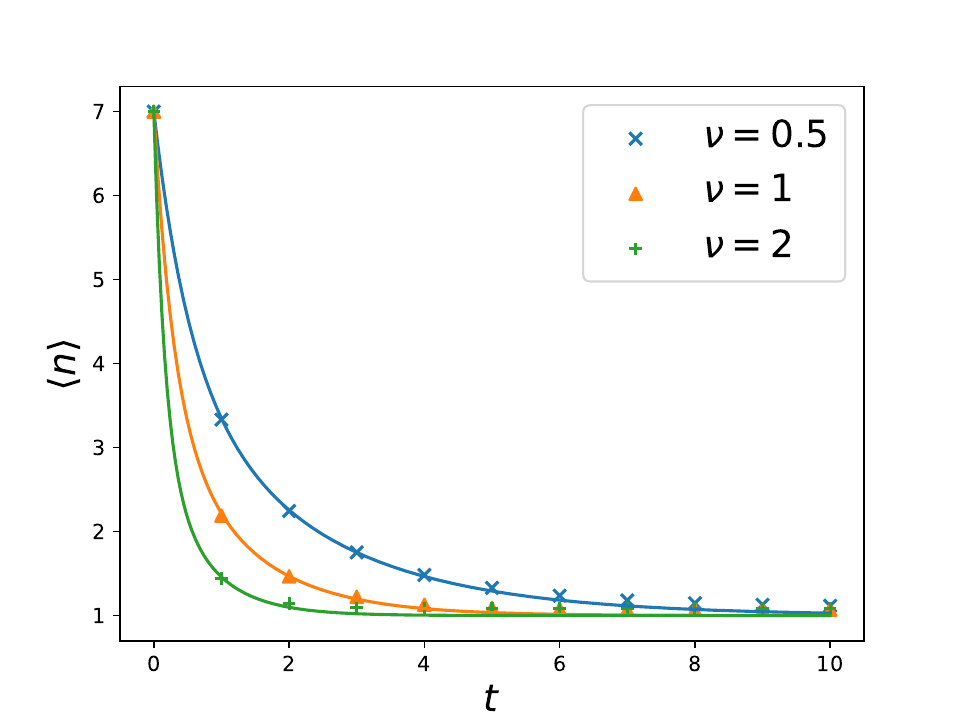}}
\caption{The evolution of the total particle number in the pair annihilation dynamics for two scenarios: (a) a system with six lattice sites, utilizing a trotter step size of $\Delta t = 1/200$, and (b) a system with seven lattice sites, employing a trotter step size of $\Delta t = 1/50$. In both cases, the simulation begins with a fully occupied initial condition and the diffusion constant $D=1$.}
\label{fig: pa particle number}
\end{figure}

In the third example, we consider the pair annihilation reaction $2A\rightarrow 0$ with a reaction rate $\nu$, alongside particles being able to freely hop on the lattice with a diffusion constant $D$. The `pseudo'-Hamiltonian governing the dynamics is given by:
\begin{equation}
\begin{aligned}
H = -\frac{1}{4}\sum_{\langle i, j\rangle}\Big[(2D+\nu)\sigma_i^x\sigma_j^x + (2D - \nu)\sigma_i^y\sigma_j^y \\ + (2D - \nu)\sigma_i^z\sigma_j^z
-i\nu \sigma_i^x\sigma_j^y - i\nu\sigma_i^y\sigma_j^x \Big] \\ + \frac{\nu}{2}\sum_i \sigma_i^z.
\end{aligned}
\end{equation}
For simplicity, we set the diffusion constant $D= 1$ in the simulation. Due to the requirement of the reaction that two particles are needed to annihilate, the total number of particles initially in the system affects the final state. We study two different cases: a fully-occupied six-site lattice and a fully-occupied seven-site lattice, as plotted in Fig.\ref{fig: pa particle number}.

In Fig.\ref{fig: pa particle number}(a), we display the total particle number $\langle n \rangle$ in the six-lattice site with a fully occupied initial condition. For different values of $\nu$, the system converges to an empty state at different rates. Fig.\ref{fig: pa particle number}(b) illustrates the time evolution of the total particle number in a seven-lattice site system with a fully occupied initial condition. Unlike the case with $6$ particles initially, the system converges to a final state with $1$ particle remaining, as it cannot find a partner to annihilate itself. In the seven-site system, the trotter step size is chosen as $\Delta t = 1/50$, and in the six-site system, the trotter step size is chosen as $\Delta t = 1/200$. Surprisingly, to achieve the same level of accuracy, we require a smaller trotter step size in the six-site system compared to the seven-site system. The challenge arises from the nature of the systems being simulated. In the case of the six-site system, it converges to a completely empty state, whereas the seven-site system converges to a superposition of different single-particle states. In such scenarios, the stationary state, representing the final outcome of the system, should correspond to one of the right eigenstates of the 'pseudo'-Hamiltonian. However, during the Trotterization process, the empty state might not be an eigenstate of most trotterized operators. Consequently, achieving an amplitude of $1$ for the fully empty state becomes difficult.

\begin{figure}
    \centering
    \includegraphics[scale=0.5]{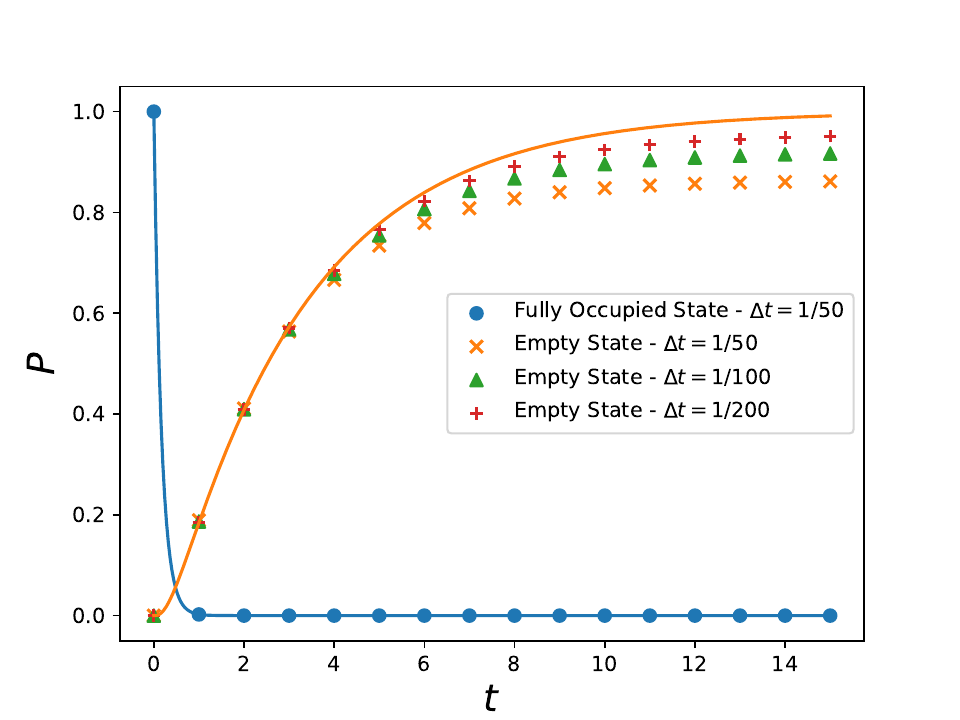}
    \caption{The evolution of the probabilities of the empty and fully occupied states in the pair annihilation dynamics with a six-lattice site system and a fully occupied initial condition: Various trotter step sizes are tested for the probability of the empty state. The solid lines depict the exact results, while the scattered markers represent the results obtained from quantum simulations. The diffusion constant $D = 1$ and the reaction rate $\nu = 0.6$.}
    \label{fig:pa probability}
\end{figure}

We further investigate the time evolution of the probabilities in the six-site system, as depicted in Fig.\ref{fig:pa probability}. In this plot, we set $\nu=1$ and vary the trotter step sizes. We plot the probabilities of the fully occupied state and the empty state, with the solid lines representing the exact results. From the plot, we observe that while the dynamics effectively capture the departure from the fully occupied state, achieving a fully empty state is challenging with the Trotter approximation. However, as we decrease the trotter step size, the simulation results gradually converge to the exact results. 

\subsection{Directed Percolation}
The last example we explore is a reaction-diffusion system that emulates the directed percolation phase transition. This system encompasses spontaneous decay and branching reactions with rates $\nu$ and $\lambda$, respectively. Additionally, particles are allowed to move freely on the lattice with a rate $D$. The `pseudo'-Hamiltonian governing this system is defined as follows:
\begin{equation}
\begin{aligned}
    H = -\frac{1}{4} \sum_{\langle i, j\rangle}\Big[
    2D \sigma_i^x\sigma_j^x + 2D\sigma_i^y\sigma_i^y + 2(D+\lambda)\sigma_i^z\sigma_j^z    \\
    +\lambda\sigma_i^z\sigma_j^x + \lambda \sigma_i^x\sigma_j^z
    +i\lambda\sigma_i^z\sigma_j^y + i\lambda\sigma_i^y\sigma_j^z
    \Big]\\
    -\frac{1}{2}\sum_i \Big[-\nu \sigma_i^z + (\lambda+\nu)\sigma_i^x + i(\lambda-\nu)\sigma_i^y\Big].
\end{aligned}
\end{equation}

In the regime where $\lambda/\nu \rightarrow 0$, the system converges to an absorbing stationary state with no particles. Conversely, as $\lambda/\nu \gg 0$, the system enters an active phase where a finite particle density is sustained. The transition between these phases, occurring for intermediate values of $\lambda/\nu$, is characterized by an active-absorbing phase transition. This transition is continuous, and its critical point falls within the directed percolation universality class.

\begin{figure}
    \centering
    \includegraphics[scale = 0.5]{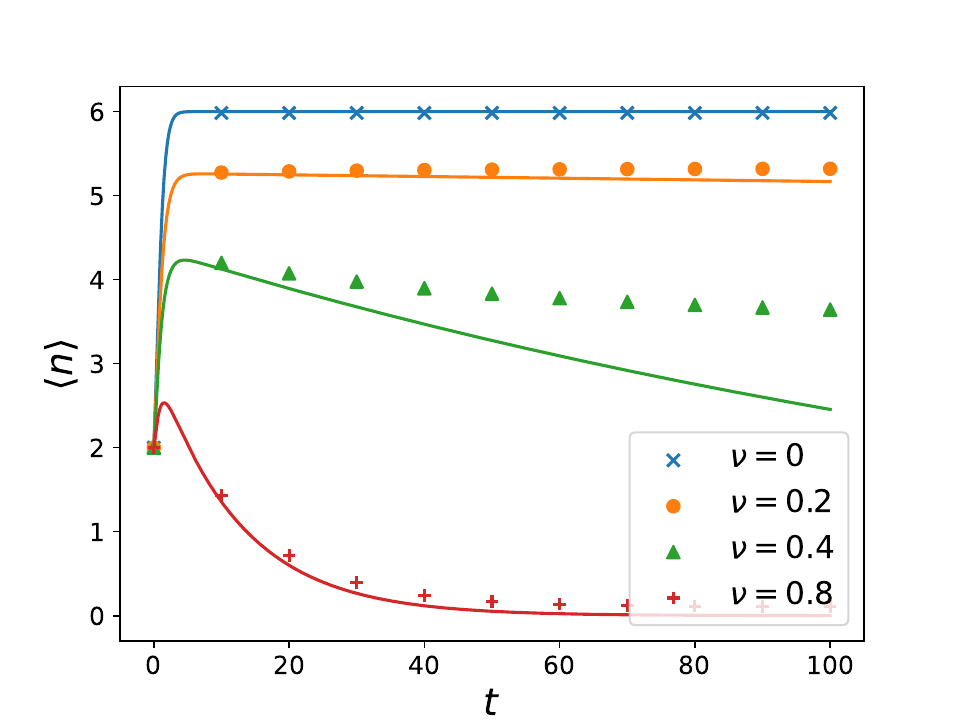}
    \caption{The evolution of particle number in the directed percolation system with six lattice sites and different decaying rates, $\nu$: The system begins with an initial condition of two occupied sites. The diffusion constant is set to $D = 1$, and the branching rate is $\lambda = 1$. The trotter step size is chosen as $\Delta t = 1/200$. The solid lines depict the exact results, while the scattered markers represent the results obtained from quantum simulations.}
    \label{fig:dp}
\end{figure}

We simulate a six-site one-dimensional lattice initially populated with two particles. The results, shown in Fig.\ref{fig:dp}, are obtained with a fixed diffusion constant $D = 1$ and branching rate $\lambda = 1$ for simplicity. The scattered markers represent the results from quantum simulation, while the solid lines represent the exact results.

As observed, when the decay rate $\nu$ is small, the system maintains a finite particle density, whereas for large decay rates, the total number of particles in the system tends towards zero. The results from the quantum simulation effectively capture the phase transition. However, due to critical slowing-down near the phase transition, it takes significantly longer to reach the stationary value, particularly evident for $\nu = 0.4$. From the limited data obtained, we can infer that the critical value $\nu_c$ lies between $0.2$ and $0.4$. It's worth noting that near the critical point, trotter errors tend to be larger compared to the results obtained deep within the two distinct phases. This discrepancy arises because deep within the phases, the system rapidly relaxes to the stationary state. While in the stationary state, the errors are mostly leakage errors, which are small. However, near the critical point, the relaxation is slow, leading to the accumulation of trotter errors. This suggests that a smaller trotter step size is needed when simulating long-time dynamics, particularly near critical points.
\section{summary and outlook}
We explore the digital quantum simulation of classical reaction-diffusion systems on lattices, focusing on the probability distribution level. To lay the groundwork, we introduce the spin representation of these systems on lattices, which provides a convenient framework for simulating them on quantum computers using qubits. We then introduce the simulation methods of implementing the dynamics. We start with trotterization, a widely-used technique for simulating time evolutions that consistently yields robust results. Next, we introduce the Pauli gadget, which provides a way to implement time evolution operator of spin systems with universal gate sets. The Pauli gadget also offers advantages in optimizing quantum circuits. Given the disparity between the master equation and Schrödinger's equation, implementing the dynamics requires handling both unitary and non-unitary operations. To address this challenge, we employ the probabilistic imaginary time evolution methods, leveraging Pauli gadgets to reduce the number of ancilla qubits required.  This method aligns well with the nature of spin representations in the system, particularly facilitating the implementation of non-unitary operations such as the time evolution operator of the form $e^{-Ht}$.
We also address the pre- and post-processing steps required to handle the different normalizations between classical probability states and quantum states. By mapping the normalization constant to a measurement of the system, we demonstrate that we can circumvent the need for exponential classical post-processing. 

Next, we apply the introduced methods to four distinct examples. Beginning with the simplest case of a single-site system involving particle generation and decay, we find that quantum simulation yields highly accurate results even with a relatively large trotter step size of $\Delta t = 1/20$. Moving on, we investigate free hopping and pair annihilation dynamics on a linear lattice, focusing on the impact of trotter errors and trotter step size. Our analysis demonstrates that sufficiently small trotter step sizes provide a reliable approximation of the exact results, both in terms of probability distributions and observables.

Finally, we explore a system exhibiting an active-absorbing phase transition. While quantum simulation accurately captures the phase transition phenomenon, we observe larger trotter errors near the critical point, necessitating the use of smaller trotter steps for improved accuracy. These comprehensive examinations showcase the versatility and efficacy of our quantum simulation approach across a range of classical reaction-diffusion systems.

One of the primary drawbacks of the method outlined in the paper stems from the combination of probabilistic imaginary time evolution (PITE) and Trotterization. As trotter errors accumulate over time, simulating longer durations necessitates smaller trotter step sizes, leading to an increased number of trotter steps. However, the probability of measuring the desired state in the PITE method decays exponentially, demanding exponentially more realizations of the simulation.

While systems away from critical points typically exhibit rapid relaxation, critical slowing-down near phase transitions result in significantly longer relaxation times. In such cases, the combined effects of Trotterization and PITE pose challenges to the accurate simulation of phase transitions. While possible methods have been proposed to address the exponential decay of success probability \cite{grover1996fast, grover2005fixed, nishi2022acceleration, chen2024reducing}, they still do not fully resolve the fundamental issue of success probability decay with an increase in trotter steps. To circumvent this challenge, one possible approach is to replace Trotterization with the Cartan decomposition \cite{kokcu2022fixed}. Although requiring classical preprocessing, the Cartan decomposition offers a way to implement time evolution operators with fixed circuit depth, thereby avoiding exponential decay while increasing the number of trotter steps. Alternatively, variational implementations of non-unitary operators \cite{jones2019variational, mcardle2019variational, motta2020determining, sun2021quantum, nishi2021implementation} can be employed to replace PITE. These methods avoid the probabilistic nature of PITE and instead rely on classical optimizations. Both approaches have the potential to yield better results for simulating dynamics over longer time scales, particularly in the study of dynamic phase transitions.

\acknowledgments
The authors express gratitude for valuable discussions with Zhangjie Qin and proofreading from Uwe C. Täuber. Additionally, the authors acknowledge the role of AI, particularly ChatGPT, in refining the language of the paper.

\newpage

\bibliography{ref}

\end{document}